\documentclass[12pt]{article}
\usepackage[centertags]{amsmath}
\usepackage{amssymb,amsfonts}
\usepackage{graphics}
\usepackage{latexsym}
\begin{document}
 
 {\LARGE \bf  A relativistic algorithm with isotropic coordinates \ \\[6mm]}
{\bf S A Ngubelanga and S D Maharaj}\\
 Astrophysics and Cosmology Research Unit\\
    School of Mathematics, Statistics and Computer Science\\
    University of KwaZulu-Natal\\
   Private Bag X54001\\
   Durban\\
   4000\\
   South Africa

\hspace{5cm}


\vspace{.5cm}

\noindent {\bf Abstract} We study spherically symmetric spacetimes 
for matter distributions with isotropic pressures. We generate new 
exact solutions to the Einstein field equations which also contains 
isotropic pressures. We develop an algorithm that produces a new 
solution if a particular solution is known. The algorithm leads to a 
nonlinear Bernoulli equation which can be integrated in terms of 
arbitrary functions. We use a conformally flat metric to show that 
the integrals may be expressed in terms of elementary functions. 
It is important to note that we utilise isotropic coordinates unlike 
other treatments.

\pagenumbering{arabic}

\section{Introduction}

\noindent We consider the interior of static perfect fluid spheres 
in general relativity with isotropic pressures. The predictions of 
general relativity have been shown to be consistent with 
observational data in relativistic astrophysics and cosmology. 
For a discussion of the physical features of a gravitating model 
we require an exact solution to the Einstein field equations. Exact 
solutions are crucial in the description of dense relativistic 
astrophysical problems. Many solutions have been found in the 
past. For some comprehensive lists of known solutions to the field 
equations refer to Delgaty and Lake \cite{Del}, Finch and Skea 
\cite{Finch}, Stephani et al. \cite{Stephani}. Many of these 
solutions are not physically reasonable. For physical reasonableness 
we require that the gravitational potentials and matter variables 
are regular, and well behaved, causality of the spacetime manifold 
is maintained and values for physical quantities, $e.g.,$ the 
mass of a dense star, are consistent with observation.

Solutions have been found in the past by making assumptions 
on the gravitational potentials, matter distribution or imposing an 
equation of state. These particular approaches do yield models 
which have interesting properties. However in principle it would be 
desirable to have a general method that produces exact solutions 
in a systematic manner. Some systematic methods generated in 
the past are those of Rahman and Visser \cite{Rahman}, Lake 
\cite{Lake}, Martin and Visser \cite{Martin}, Boonserm et al. 
\cite{Boonserm}, Herrera et al. \cite{Herrera}, Chaisi and 
Maharaj \cite{Chaisi1} and Maharaj and Chaisi \cite{Chaisi2}. In 
general relativity we have the freedom of using any well defined 
coordinate system. The references mentioned above mainly use 
canonical coordinates. The use of isotropic coordinates may 
provide new insights and possibly lead to new solutions. This is 
the approach that we follow in this paper. We generate a new 
algorithm producing a new solution to Einstein field equations in 
isotropic coordinates. From a given solution we can find a new 
solution with isotropic pressures.

The object of this paper is to find new classes of exact solutions 
of the Einstein field equations with an uncharged isotropic matter 
distribution from a given seed metric. In Section 2, we derive the 
Einstein field equations for neutral perfect fluids in static spherically 
symmetric spacetime. We introduce new variables due to 
Kustaanheimo and Qvist \cite{Kustaano} to rewrite the field equations 
and the condition of pressure isotropy in equivalent forms. In Section 3, 
we introduce our algorithm and the master nonlinear second order 
differential equation containing two arbitrary functions, that has to 
be solved. In Section 4, we present new classes of exact solutions in 
terms of the arbitrary functions. In Section 5, we give an example for 
a conformally flat metric showing that the integrals generated in Section 
4 may be explicitly evaluated. In Section 6, we summarise the results 
obtained in this paper.

\section{The model}

We are modelling the interior of a dense relativistic star in strong gravitational 
fields. The line element of the interior spacetime, with isotropic coordinates, 
has the following form

\begin{equation}
\label{eq:g1}  ds^{2} = -A^{2}(r)dt^{2} + B^{2}(r)[dr^{2} + r^{2} (d \theta^{2} + \sin^{2} \theta d \phi^2)]
\end{equation}

\noindent where $A(r)$ and $B(r)$ are arbitrary functions representing the 
gravitational potentials. Relativistic compact objects such as neutron stars 
 in astrophysics are described by this line element. The energy momentum 
tensor for the interior of the star has the form of a perfect fluid 

\begin{equation}
\label{eq:g2}  	T^{ab} = (\rho + p)u^a u^b + p g^{ab}
\end{equation}

\noindent where $\rho$ is the energy density and $p$ is the isotropic pressure. 
These quantities are measured relative to a timelike unit four-velocity 
$u^{a}$ ($u^a u_{a}=-1$).

\vspace{1cm}
The Einstein field equations for (\ref{eq:g1}) and (\ref{eq:g2}) have the form 

\begin{subequations} \label{eq:g3}
\begin{eqnarray}
\label{eq:g3a} \rho &=& -\frac{1}{B^{2}}\left[2 \frac{B''}{B}-\frac{B'}{B}\left(\frac{B'}{B}-\frac{4}{r}\right)\right]\\		\nonumber\\ 
\label{eq:g3b}    p &=& 2 \frac{A'}{A}\left(\frac{B'}{B^{3}}+ \frac{1}{r} \frac{1}{B^{2}}\right)+
                         \frac{B'}{B^{3}}\left(\frac{B'}{B}+ \frac{2}{r}\right)\\									\nonumber\\   
\label{eq:g3c}   p &=& \frac{1}{B^{2}}\left(\frac{A''}{A} + \frac{1}{r} \frac{A'}{A}\right)+
                         \frac{1}{B^{2}}\left[\frac{B''}{B}-\frac{B'}{B}\left(\frac{B'}{B}-\frac{1}{r}\right)\right]
\end{eqnarray}
\end{subequations}

\noindent in isotropic coordinates. Primes denote differentiation with 
respect to the radial coordinate $r$. On equating  (\ref{eq:g3b}) and 
(\ref{eq:g3c}) we obtain the condition of pressure isotropy which has 
the form: 

\begin{equation}
\label{eq:g4} \frac{A''}{A}+\frac{B''}{B}=\left(\frac{A'}{A}+\frac{B'}{B}\right)\left(2\frac{B'}{B}+\frac{1}{r}\right)
\end{equation}
\noindent This is the master equation which has to be integrated 
to produce an exact solution to the field equations.

\vspace{1cm}
It is possible to write the system  (\ref{eq:g3}) in an equivalent form 
by introducing new variables. We utilize a transformation that has 
proven to be helpful in relativistic stellar physics. We introduce the 
new variables 

\begin{equation}
\label{eq:g5}      x \equiv  r^{2},       \hspace{2.5cm}       L  \equiv B^{-1},     \hspace{2.5cm}  G \equiv  LA
\end{equation}

\noindent The above transformation was first suggested by 
Kustaanheimo and Qvist \cite{Kustaano}. On applying transformation 
(\ref{eq:g5}) in the field equations (\ref{eq:g3}) we obtain the 
equivalent system 

\begin{subequations} \label{eq:g6}
\begin{eqnarray}
\label{eq:g6a} \rho        &=& 4[2xLL_{xx}-3(xL_{x}-L)L_{x}]\\                                                       \nonumber\\
\label{eq:g6b}  p           &=& 4L(L-2xL_{x})\frac{G_{x}}{G}-4(2L-3xL_{x})L_{x}\\                                      \nonumber\\
\label{eq:g6c}  p           &=& 4xL^{2}\frac{G_{xx}}{G}+4L(L-2xL_{x})\frac{G_{x}}{G}-4(2L-3xL_{x})L_{x}-8xLL_{xx}
\end{eqnarray}
\end{subequations}

\noindent We note that the above equations (\ref{eq:g6}) are highly 
nonlinear in both $L$ and $G$. In this system there are three independent 
equations and four unknowns $\rho$, $p$, $L$ and $G$. So we need to 
choose the functional form for $L$ or $G$ in order to integrate and obtain 
an exact solution. The value of the transformation (\ref{eq:g5}) is 
highlighted in the reduction of the condition of pressure isotropy. On 
equating equations (\ref{eq:g6b}) and (\ref{eq:g6c}) we get 

\begin{equation}
\label{eq:g7}   LG_{xx} = 2GL_{xx}
\end{equation}

\noindent which is the new condition of pressure isotropy which has 
a simpler compact form.

\section{The algorithm}

\noindent It is possible to find new solutions to the Einstein's 
equations from a given seed metric. Examples of this process 
are given in the treatments of Chaisi and Maharaj \cite{Chaisi1} 
and Maharaj and Chaisi \cite{Chaisi2}. They found new models, 
with anisotropic pressures, from a given seed isotropic metric in 
Schwarzschild coordinates. Our intention is to find new models, 
with isotropic pressures, from a given solution in terms of the 
isotropic line element (\ref{eq:g1}).

 We can provide some new classes of exact solutions to the 
Einstein field equations by generating a new algorithm that produces 
a model from a given solution. We assume a known solution of the 
form $(\bar{L}, \bar{G})$ so that 

\begin{equation}
\label{eq:g8}      \bar{L} \bar{G}_{xx} = 2 \bar{G} \bar{L}_{xx}
\end{equation}

\noindent holds. We seek a new solution $(L,G)$ given by 
 
\begin{equation}
\label{eq:g9}     L = \bar{L} e^{g(x)},           \hspace{2.5cm}     G =\bar{G} e^{f(x)}
\end{equation}
 
\noindent where $f(x)$ and $g(x)$ are arbitrary functions. On 
substituting equation (\ref{eq:g9}) into (\ref{eq:g7}) we obtain 

\begin{eqnarray}
\label{eq:g10}			(\bar{L}\bar{G}_{xx}-2\bar{G}\bar{L}_{xx})+2(\bar{L}\bar{G}_{x}f_{x}-2\bar{G}\bar{L}_{x}g_{x})+
				\bar{L}\bar{G}(f_{xx}-2g_{xx})+\bar{L}\bar{G}(f^{2}_{x}-2g^{2}_{x})=0
\end{eqnarray}

\noindent which is given in terms of two arbitrary functions $f(x)$ and 
$g(x)$. Then realizing that $(\bar{L}, \bar{G})$ is a solution of 
(\ref{eq:g7}) and using (\ref{eq:g8}) we obtain the reduced result

\begin{equation}
\label{eq:g11}      (f_{xx}-2g_{xx})+2\left(\frac{\bar{G}_{x}}{\bar{G}}f_{x}-2\frac{\bar{L}_{x}}{\bar{L}}g_{x}\right)+(f^2_{x}-2g^2_{x})=0
\end{equation}

\noindent We need to demonstrate the existence of functions 
$f(x)$ and $g(x)$ that satisfy (\ref{eq:g11}). In general it is 
difficult to integrate equation (\ref{eq:g11}), since it is given 
in terms of two arbitrary functions which are nonlinear.

\section{New solutions}

\noindent We consider several cases of equation (\ref{eq:g11}) 
for which we have been able to complete the integration.

\subsection{ $g(x)$ is specified}

We can integrate (\ref{eq:g11}) if $g(x)$ is specified. As a simple 
example we take $g(x) = 1$. Then (\ref{eq:g11}) becomes 

\begin{equation}
\label{eq:g12}      f_{xx}+2\frac{\bar{G}_{x}}{\bar{G}}f_{x}+f^2_{x}=0
\end{equation}

\noindent which is nonlinear in $f$. This is a first order Bernoulli 
equation in $f_{x}$. We can rewrite (\ref{eq:g12}) in the form 

\begin{equation}
\label{eq:g13}   \left(\frac{1}{f_{x}}\right)_{x} - 2\left(\frac{\bar{G}_{x}}{\bar{G}}\right) \left(\frac{1}{f_{x}}\right) = 1
\end{equation} 

\noindent It is possible to integrate equation (\ref{eq:g13}) since it is linear in 
$\frac{1}{f_{x}}$ to obtain 

\begin{equation}
\label{eq:g14}        f_{x}= \bar{G}^{-2}\left(\int{\bar{G}^{-2}dx +c_{1}}\right)^{-1}
\end{equation}
 
\noindent We can formally integrate (\ref{eq:g14}) to obtain the 
function $f(x)$ as  

\begin{equation}
\label{eq:g15}     f(x) = \int{\left[\bar{G}^{-2}\left(\int{\bar{G}^{-2} dx +c_{1}}\right)^{-1}\right]}dx+c_{2}
\end{equation}

\noindent where $c_{1}$ and $c_{2}$ are arbitrary constants.

Then the new solution to equation (\ref{eq:g7}) has the form 

\begin{subequations}    \label{eq:g16}
\begin{eqnarray}
\label{eq:g16a}        L &=& \bar{L}                                        \\          \nonumber\\
\label{eq:g16b}        G &=& \bar{G} \exp{\left( \int{\left[\bar{G}^{-2}\left(\int{\bar{G}^{-2} dx +c_{1}}\right)^{-1}\right]}dx+c_{2}\right)}
\end{eqnarray}
\end{subequations}

\noindent Therefore we have shown that if a solution $(\bar{L}, \bar{G})$ 
to the field equations is known, then a new solution $(L, G)$ is given by 
(\ref{eq:g16}).

\subsection{ $f(x)$ is specified}

We can also integrate  (\ref{eq:g11}) if $f(x)$ is specified. 
As another simple example we take $f(x) = 1$. Then  equation 
(\ref{eq:g11}) becomes 

\begin{equation}
\label{eq:g17}      g_{xx}+2\frac{\bar{L}_{x}}{\bar{L}}g_{x}+g^2_{x}=0
\end{equation}

\noindent which is nonlinear in $g$. This  is a first order Bernoulli equation
 in $g_{x}$. The differential equation (\ref{eq:g17}) has a form similar to 
(\ref{eq:g12}) in section $4.1$. Following the same procedure we obtain 

\begin{equation}
\label{eq:g18}     g(x) = \int{\left[\bar{L}^{-2}\left(\int{\bar{L}^{-2} dx +c_{1}}\right)^{-1}\right]}dx+c_{2}
\end{equation}

\noindent where $c_{1}$ and $c_{2}$ are arbitrary constants. 

Then another new solution to equation (\ref{eq:g7}) is given by 

\begin{subequations}    \label{eq:g19}
\begin{eqnarray}
\label{eq:g19a}        G &=& \bar{G}                                         \\          \nonumber\\
\label{eq:g19b}          L &=& \bar{L} \exp{\left( \int{\left[\bar{L}^{-2}\left(\int{\bar{L}^{-2} dx +c_{1}}\right)^{-1}\right]}dx+c_{2}\right)}
\end{eqnarray}
\end{subequations}

\noindent Therefore we have determined that if a solution $(\bar{L}, \bar{G})$ 
to the field equations is known then a new solution $(L, G)$ is given by 
(\ref{eq:g19}). Note that the solution (\ref{eq:g19}) is different from that 
presented in (\ref{eq:g16}).

\subsection{ $g(x)=\alpha f(x)$}

\vspace{0.5cm}

\noindent We can integrate (\ref{eq:g11})  if a relationship between the 
functions $f(x)$ and $g(x)$ exists. We illustrate this feature by assuming
that
\begin{equation}
 \label{eq:g20}                    g(x) = \alpha f(x)
\end{equation}

\noindent where $\alpha$ is an arbitrary constant. Then (\ref{eq:g11}) 
becomes

\begin{equation}
\label{eq:g21}  f_{xx} + \frac{2}{1-2\alpha} \left(\frac{\bar{G_{x}}}{\bar{G}}-2\alpha \frac{\bar{L}_{x}}{\bar{L}}\right)f_{x} + \left(\frac{1-2\alpha^2}{1-2\alpha}\right) f^{2}_{x} = 0
\end{equation}

\noindent which is a first order Bernoulli equation in $f_{x}$. For 
convenience we let 

\begin{equation}
\label{eq:g22}                 \Theta = \left(\frac{1-2\alpha^2}{1-2\alpha}\right), \hspace{1cm} \eta =  \frac{2}{1-2\alpha}          , \hspace{1cm} \alpha \neq \frac{1}{2}
\end{equation}

\noindent so that we can write (\ref{eq:g21}) as

\begin{equation}
\label{eq:g23} \left(\frac{1}{f_{x}}\right)_{x} -\eta \left(\frac{\bar{G_{x}}}{\bar{G}}-2\alpha \frac{\bar{L}_{x}}{\bar{L}}\right)\left(\frac{1}{f_{x}}\right)=\Theta
\end{equation}

\noindent which is linear in $\frac{1}{f_{x}}$. We integrate (\ref{eq:g23}) to obtain 

\begin{equation}
\label{eq:g24}     f_{x} =  \left(\frac{\bar{L}^{2\alpha}}{\bar{G}}\right)^{\eta}\left[\Theta \int{\left(\frac{\bar{L}^{2\alpha}}{\bar{G}}\right)^{\eta}dx}+c_{1}\right]^{-1}
\end{equation}

\noindent We now formally integrate (\ref{eq:g24}) to obtain

\begin{equation}
\label{eq:g25}     f(x) = \int{\left(\left(\frac{\bar{L}^{2\alpha}}{\bar{G}}\right)^{\eta}\left[ \Theta\int{\left(\frac{\bar{L}^{2\alpha}}{\bar{G}}\right)^{\eta}dx}+c_{1}\right]^{-1}\right)}dx+c_{2}
\end{equation}

\noindent where $c_{1}$ and $c_{2}$ are constants.

\vspace{0.5 cm}
We now have a new solution of (\ref{eq:g7}) given by 

\begin{subequations}        \label{eq:g26}
\begin{eqnarray}
\label{eq:g26a}    L &=& \bar{L} \exp{\alpha \left[\int{\left(\left(\frac{\bar{L}^{2\alpha}}{\bar{G}}\right)^{\eta}\left[ \Theta \int{\left(\frac{\bar{L}^{2\alpha}}{\bar{G}}\right)^{\eta}dx}+c_{1}\right]^{-1}\right)}dx+c_{2}\right]}\\         \nonumber\\
\label{eq:g26b}    G &=& \bar{G} \exp{ \left[\int{\left(\left(\frac{\bar{L}^{2\alpha}}{\bar{G}}\right)^{\eta}\left[\Theta\int{\left(\frac{\bar{L}^{2\alpha}}{\bar{G}}\right)^{\eta}dx}+c_{1}\right]^{-1}\right)}dx+c_{2}\right]}
\end{eqnarray}
\end{subequations}

\noindent where $\Theta$ and $\eta$ are given in (\ref{eq:g22}). 
Therefore we have demonstrated that if a solution $(\bar{L}, \bar{G})$ 
to the field equations is specified then a new solution $(L,G)$ is provided by 
(\ref{eq:g26}).

Some special cases related to (\ref{eq:g26}) should be pointed out. 
These relate to $\alpha = 1$,  $ \pm \frac{1}{\sqrt{2}}$,  
$\frac{1}{2}$. We consider each in turn.

\vspace{1cm}
{\bf Case $(i)$:} $\alpha = 1$

\noindent With $\alpha = 1$ we find that (\ref{eq:g26}) becomes

\begin{subequations}    \label{eq:g27}
\begin{eqnarray}
\label{eq:g27a}        L &=& \bar{L} \exp{\left(\int{\left[\frac{\bar{G}^{2}}{\bar{L}^{4}}\left(\int{\frac{\bar{G}^{2}}{\bar{L}^{4}} dx +c_{1}}\right)^{-1}\right]}dx+c_{2}\right)}\\          \nonumber\\
\label{eq:g27b}        G &=& \bar{G} \exp{\left(\int{\left[\frac{\bar{G}^{2}}{\bar{L}^{4}}\left(\int{\frac{\bar{G}^{2}}{\bar{L}^{4}} dx +c_{1}}\right)^{-1}\right]}dx+c_{2}\right)}
\end{eqnarray}
\end{subequations}

\noindent which is a simple form.

\vspace{1cm}
{\bf Case $(ii)$:} $\alpha = \pm \frac{1}{\sqrt{2}}$

\noindent If we set $\alpha = \pm \frac{1}{\sqrt{2}}$ then (\ref{eq:g26}) becomes

\begin{subequations}        \label{eq:g28}
\begin{eqnarray}
\label{eq:g28a}    L &=& \bar{L} \exp{\left[\pm \frac{1}{\sqrt{2}} \left(c_{1}\int{\left(\frac{\bar{L}^{\pm \sqrt{2}}}{\bar{G}}\right)^{\frac{2}{1-(\pm \sqrt{2})}}} dx + c_{2}\right)\right]}\\        \nonumber\\
\label{eq:g28b}    G &=& \bar{G} \exp{\left[c_{1}\int{\left(\frac{\bar{L}^{\pm \sqrt{2}}}{\bar{G}}\right)^{\frac{2}{1-(\pm \sqrt{2})}}} dx + c_{2}\right]}
\end{eqnarray}
\end{subequations}

\noindent which is another simple case.

\vspace{1cm}
{\bf Case $(iii)$:} $\alpha = \frac{1}{2}$

\noindent If $\alpha = \frac{1}{2}$ then (\ref{eq:g26}) is not valid. For this 
case,  equation (\ref{eq:g11}) becomes 

\begin{equation}
\label{eq:g29} 	f_{x}\left[f_{x}+4\left(\frac{\bar{G}_{x}}{\bar{G}}-\frac{\bar{L}_{x}}{\bar{L}}\right)\right]=0
\end{equation}

\noindent When $f$ is constant then $g$ is also constant by 
(\ref{eq:g20}); then (\ref{eq:g7}) does not produce a new 
solution because of (\ref{eq:g9}). When $f$ is not constant 
then we can integrate (\ref{eq:g29}) to produce the solution 

\begin{subequations}    \label{eq:g30}
\begin{eqnarray}
\label{eq:g30a}        L &=& K    \frac{\bar{L}^3}{\bar{G}^2} \\          \nonumber\\
\label{eq:g30b}        G &=& K    \frac{\bar{L}^4}{\bar{G}^3} 
\end{eqnarray}
\end{subequations}

\noindent where $K$ is a constant. Thus $\alpha = \frac{1}{2}$ 
generates another new solution $(L,G)$ to (\ref{eq:g11}).

\section{Example}

We show by means of a specific example that the integrals 
generated in section 4 may be evaluated to produce a new 
exact solution to the field equations in terms of elementary 
functions. In our example we choose 

\begin{subequations}    \label{eq:g31}
\begin{eqnarray}
\label{eq:g31a}			{\bar{L}} &=&b+ax\\		\nonumber\\
\label{eq:g31b}		  	{\bar{G}} &=&1+cx
\end{eqnarray}
\end{subequations}

\noindent Then the corresponding line element is given by 

\begin{eqnarray}
\label{eq:g32}		ds^2 = -\left(\frac{1+cr^2}{b+ar^2}\right)^2 dt^2+\left(\frac{1}{b+ar^2}\right)^2 (dr^2+r^2(d\theta^2+\sin^2 \theta d\phi^2)
\end{eqnarray}

\noindent which is conformally flat. The energy density
for the metric (\ref{eq:g32}) is constant so that we 
have the Schwarzschild interior solution in isotropic 
coordinates.

Conformally flat metrics are important in gravitational 
physics in a general relativistic setting. They arise, for 
instance in the gravitational collapse of a radiating star, 
as shown in the treatments of Herrera et al. 
\cite{Herrera1}, Maharaj and Govender \cite{Maharaj}, 
Misthry et al. \cite{Misthry} and Abebe et al. 
\cite{Abebe}. For the choice of (\ref{eq:g31}) we find 
that (\ref{eq:g27}) becomes 

\begin{subequations}    \label{eq:g33}
\begin{eqnarray}
\label{eq:g33a}        L &=& (b+ax) \exp{\left(\int{\left[\frac{(1+cx)^{2}}{(b+ax)^{4}}\left(\int{\frac{(1+cx)^{2}}{(b+ax)^{4}} dx +c_{1}}\right)^{-1}\right]}dx+c_{2}\right)}\\          \nonumber\\
\label{eq:g33b}        G &=& (1+cx)\exp{\left(\int{\left[\frac{(1+cx)^{2}}{(b+ax)^{4}}\left(\int{\frac{(1+cx)^{2}}{(b+ax)^{4}} dx +c_{1}}\right)^{-1}\right]}dx+c_{2}\right)}
\end{eqnarray}
\end{subequations}

\noindent The integrals in (\ref{eq:g33}) can be evaluated 
and we obtain 

\begin{subequations}    \label{eq:g34}
\begin{eqnarray}
\label{eq:g34a}			 L   &=& \frac{1}{(b+ax)^2} U(x)\\			\nonumber\\
\label{eq:g34b}			 G  &=& \frac{(1+cx)}{(b+ax)^3} U(x)
\end{eqnarray}
\end{subequations}

\noindent where $c_{1}=0$ and $c_{2} = 1$ and 
we have set

\begin{equation}
\label{eq:g35}			 U(x) = b^2c^2+abc(1+3cx)+ a^2(1+3cx+3c^2x^2) 
\end{equation}

\noindent  Thus the known solution $(\bar{L},\bar{G})$ in 
(\ref{eq:g31}) produces a new solution $(L,G)$ in 
(\ref{eq:g34}). The line element for the new 
solution has the form 

\begin{equation}
\label{eq:g36}	  	ds^2 = -\left(\frac{1+cr^2}{b+ar^2}\right)^2 dt^2+\left(\frac{(b+ar^2)^2}{U(r)}\right)^2(dr^2+r^2(d\theta^2+\sin^2 \theta d\phi^2))
\end{equation}

\noindent where $U(r)$ is given by (\ref{eq:g35}). Thus our algorithm has produced a new (not conformally flat) 
solution to the Einstein's field equations from a seed conformally flat model.

\section{Conclusion}

\begin{figure}[h]
    \centering
    \includegraphics{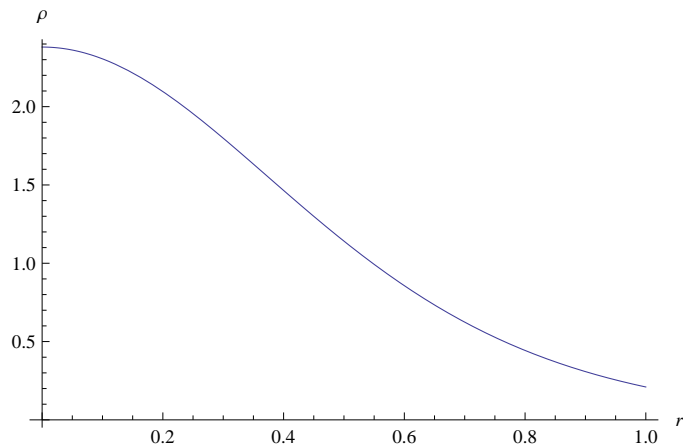}
    \caption{Energy density $\rho$}
 \label{Figure1}
\end{figure}

\begin{figure}[h]
   \centering
   \includegraphics{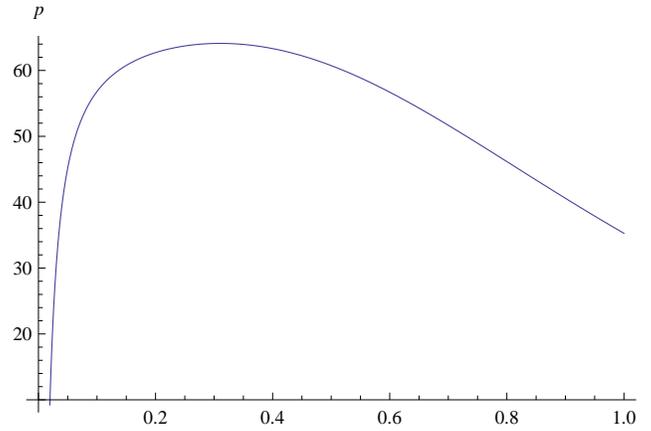}
 \caption{Pressure $p$}
\label{Figure2}
\end{figure}

\begin{figure}[h]
   \centering
  \includegraphics{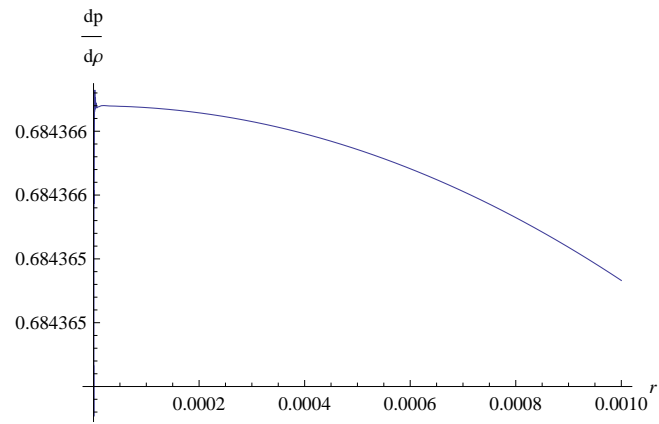}
  \caption{Speed of sound $\frac{dp}{d\rho}$}
 \label{Figure3}
\end{figure}

\noindent We now comment on the physical properties of the example.
We have generated plots for the energy density $\rho$, pressure $p$, 
and the speed of sound  in Figures \ref{Figure1}-\ref{Figure3}, respectively. These 
graphical plots indicate that $\rho$ and $p$ are positive and well behaved. The speed of sound is less than the speed
of light as required for causality. Therefore the 
algorithm presented in this paper produces new solutions which are 
physically reasonable.

We have generated an algorithm to produce a new solution 
to the Eistein field equations from a given seed metric. We observe 
that the resulting model contains isotropic pressures unlike the approach 
of Chaisi and Maharaj \cite{Chaisi1} and Maharaj and Chaisi 
\cite{Chaisi2}; in their treatment the new model has anisotropic 
pressures. Another advantage of our approach is the use of isotropic 
coordinates in the formulation of the condition of pressure isotropy. 
This may leads to new insights into the behaviour of gravity since 
previous treatments mainly utilised canonical coordinates. The 
algorithm produced a new solution in terms of integrals containing 
arbitrary functions. We have shown, with the help of a conformally 
flat metric, that these integrals may be evaluated in terms of 
elementary functions. This example suggests that our approach 
may be extended to other physically relevant metrics.

\vspace{1cm}

\noindent {\bf Acknowledgements}\\
SAN thanks the National Research Foundation and the University of KwaZulu-Natal
for financial support. SDM acknowledges that this work is based upon research 
supported by the South African Research Chair Initiative of the Department of 
Science and Technology and the National Research Foundation.\\

\end{document}